# Autofocusing Self-Imaging: The Symmetric Pearcey Talbot-like Effect *

**Jiajia Zhao, You Wu, Zejia Lin, Danlin Xu, Haiqi Huang, Zhifeng Tu, Hongzhan Liu, Lingling Shui and Dongmei Deng**
Guangdong Provincial Key Laboratory of Nanophotonic Functional Materials and Devices
Guangzhou 510631
China
School of Information and Optoelectronic Science and Engineering, South China Normal University
South China Normal University
Guangzhou 510006
China
{Lingling Shui}shuill@m.scnu.edu.cn
{Dongmei Deng}dengdongmei@m.scnu.edu.cn
{Jiajia Zhao, You Wu}These authors contribute equally

**Chuangjie Xu**
Room 832, Apartment 353, Sun Yat-Sen University
Guangzhou 510275
China

## Abstract

The Talbot-like effect of symmetric Pearcey beams (SPBs) is presented numerically and experimentally in the free space. Owing to the Talbot-like effect, the SPBs have the property of periodic and multiple autofocusing. Meanwhile, the focal positions and focal times of SPBs are controlled by the beam shift factor and the distribution factors. What's more, the beam shift factor can also affect the Talbot-like effect and the Talbot period. Therefore, several tiny optical bottles can be generated under the appropriate parameter setting. It is believed that the results can diversify the application of the Talbot effect.

**Keywords** Talbot-like effect, · Symmetric Pearcey beams · Self-imaging

## 1  Introduction

The Talbot effect, first discovered by Henry Fox Talbot [1] in 1836, was also called lens-less imaging or self-imaging. About half a century later, Lord Rayleigh explained this phenomenon as the interference of the diffracted beams and performed the analytical calculations for the plane wave illumination [2]. He found the exact reconstructed images of the grating repeated with the longitudinal period, Talbot period $Z = 2m \frac{d^2}{\lambda}$ where $d$ is the grating constant, $\lambda$ denotes the wavelength, and $m$ is a positive integer. In 1996, Montgomery proved that lateral periodicity of the object is a sufficient condition for the achievement of its reconstruction [3]. It means we can obtain the periodic reconstruction of the wave without using optical lenses when the Montgomery condition is satisfied. Therefore, Kyva1sk pointed out in Ref. [4], the longitudinal periodicity of the wavefields can be presented as a superposition of nondiffracting beams or modes. Then in Ref. [5], similar conclusions are drawn in the frequency and the time domains, respectively. Many scientists had studied the Talbot effect and generalized it to practical applications, such as imaging processing and synthesis [6], photolithography [7], optical testing [8], optical metrology [9], etc. This lens-less imaging phenomenon has also been observed in various areas of research, such as atomic optics [10], nonlinear optics [11], quantum optics [12], waveguide arrays [13], plasmon [14], photonic lattices [15], metamaterials [16] and the fractional Schrödinger systems [17], etc. Some corresponding conceptual extensions of the Talbot effect have been proposed, including the quantum Talbot effect [12], nonlinear Talbot effect [11], parity-time symmetric Talbot effect [15], plasmon Talbot effect [14], Talbot-Lau effect [18], and Airy-Talbot effect [5], etc. On the other hand, the autofocusing beams have recently attached great interest, and different autofocusing beams have been produced [19, 20, 21]. We know that symmetric Pearcey beams (SPBs) have excellent properties such as autofocusing, nondiffraction, vortex-guiding [21] and Pearcey beams have periodic focus and dispersion in parabolic refractive index [29]. Meanwhile, the Talbot effect of some autofocusing beams has been studied [5, 22, 23, 24] and the Talbot effect has a wide range of applications in optics [30, 31, 32]. What will happen if we combine SPBs with the Talbot effect, making SPBs periodic in free space? Therefore, we study the self-imaging of the SPBs by taking the spectrum of SPBs as a periodic function, which generates a Talbot-like field [25], and we call it the symmetric Pearcey Talbot-like effect. We add the Gaussian term to SPTBs, which can be implemented in an actual experiment. Furthermore, we add a cubic phase to observe the change of the optical field more clearly. At last, we explore the influence of parameter setting on the optical field of the Talbot-like effect. We are pleasantly surprised to find that the symmetric Pearcey Talbot-like effect can form several tiny optical bottles (OBs) under the appropriate beam shift factor.

## 2 Theory model

We know that the Pearcey integral is defined as [27]: $Pe(X, Y) = \int_{-\infty}^{+\infty} \exp[i(s^4 + Ys^2 + Xs)]ds$, which can be calculated numerically using a contour rotation in the complex plane [28]. Under the paraxial system, we can obtain the spatial spectrum of SPBs [21]:

$$\widetilde{SPB}(k_x, k_y) = \widetilde{Pe}(k_x bw_0, p)\widetilde{Pe}(k_y bw_0, p) \tag{1}$$

where $k_x$ and $k_y$ are the spatial frequencies of $x$ and $y$, respectively, $b$ and $p$ indicate the distribution factors, $w_0$ is the beam waist and $\widetilde{Pe}(\cdot)$ is the spatial spectrum of the Pearcey function. Meanwhile, the propagation of the spatial beam in the free space along the z-axis derives from the (2+1) dimensional Schrödinger equation: $2ik\frac{\partial \psi}{\partial z} + \frac{\partial^2 \psi}{\partial x^2} + \frac{\partial^2 \psi}{\partial y^2} = 0$, where $\psi$ is an optical field, $k = 2\pi/\lambda$ is the wavenumber, and $\lambda$ denotes the wavelength.

A superposition of nondiffracting beams will present the longitudinal periodicity of the Talbot-like effect [4, 25]. There are two methods to demonstrate the Talbot-like effect for the one dimensional paraxial wave equation in the free space. The first method is the moving frame approach, which derives from the free space[22]. The second method is to look at a field whose spatial spectrum is a periodic function. The two methods are equivalent, and both prove that the superposition of several single beams with constant beam shift can form the Talbot-like effect of beams [5]. We choose the second method, modifying the spatial spectrum of SPBs. A two dimensional periodic spatial spectrum $f(k_x, k_y)$ can be defined as [5]

$$f(k_x, k_y) = f(k_x + v_n, k_y + v_n) = f\left(k_x + \frac{2\pi n}{\triangle}, k_y + \frac{2\pi n}{\triangle}\right) = \sum_n c_n \exp(-ik_x n\triangle)\exp(-ik_y n\triangle), \tag{2}$$

where $c_n$ is a superposition coefficient, $v_n = \frac{1}{\triangle}$ is the spatial frequency of period ($\triangle$ is the beam shift factor) and $n$ is a positive integer representing the number of superimposed beams. Now, we multiply the periodic spatial spectrum Eq. (2) by the spatial spectrum of SPBs Eq. (1). Inverse Fourier transforming it [5], we can obtain the symmetric Pearcey Talbot-like beams (SPTBs):

$$SPTB(x, y) = F^{-1}\left[\widetilde{SPB}(k_x, k_y)f(k_x, k_y)\right]$$

$$= F^{-1}\left[\widetilde{Pe}(k_x bw_0, p)\widetilde{Pe}(k_y bw_0, p)\sum_n c_n \exp(-ik_x n\triangle)(-ik_y n\triangle)\right]$$

$$= F^{-1}[2\pi\exp(ik_x^4 w_0^4 b^4 + ipk_x^2 w_0^2 b^2)\sum_n c_n \exp(-ik_x n\triangle)\times 2\pi\exp(ik_y^4 w_0^4 b^4 + ipk_y^2 w_0^2 b^2)\sum_n c_n \exp(-ik_y n\triangle)]$$

$$= \sum_n c_n Pe\left(\frac{x-\triangle n}{bw_0}, p\right)Pe\left(\frac{y-\triangle n}{bw_0}, p\right),$$

(3)where $F^{-1}$ indicates the inverse Fourier transform and $Pe(\cdot)$ represents the Pearcey function. To get a physically finite energy beam, we add the Gaussian term $Ga(x,y) = \sum_n c_n \exp[-\frac{(x-\triangle n)^2 + (y-\triangle n)^2}{w_0^2}]$ to SPTBs. So we can generate symmetric Pearcey Gaussian Talbot-like beams (SPGTBs) in an actual experiment. The initial field of the SPGTBs can be expressed as:

$$SPGTB(x, y) = SPTB(x, y)Ga(x, y) = \sum_n c_n Pe\left(\frac{x-\triangle n}{bw_0}, p\right)Pe\left(\frac{y-\triangle n}{bw_0}, p\right)\times\exp[-\frac{(x-\triangle n)^2 + (y-\triangle n)^2}{w_0^2}]. \tag{4}$$

Similarly, we superimpose on the SPGTBs a cubic phase factor $\exp[i\,\beta\frac{(x-\triangle n)^{\frac{3}{2}}(y-\triangle n)^{3}}{w_0^3}]$, giving SPGTBs a parabolic trajectory. The initial field of the cubic symmetric Pearcey Gaussian Talbot-like beams (CSPGTBs) can be expressed as:

$$\text{CSPGTB}(x,y) = \sum_n c_n \text{Pe}\left(\frac{x-\triangle n}{bw_0}, p\right) \text{Pe}\left(\frac{y-\triangle n}{bw_0}, p\right) \times \exp[-\frac{(x-\triangle n)^2 + (y-\triangle n)^2}{w_0^2}] \times \exp[i\,\beta\frac{(x-\triangle n)^3 + (y-\triangle n)^3}{w_0^3}],$$

(5)

where $\beta$ represents the coefficient of the cubic phase, which can control the curvature of the trajectory. Other parameters are the same as those in Eq. (4). We get the numerical simulation by the split-step Fourier transform method. This paper assumes that $w_0 = 1.5mm$, $\triangle = 0.51mm$, $n = [-5, 5]$, and $c_n = [\cdot \bullet \cdot, 1, 1, 1, \cdot \bullet \cdot]$ without special instructions, that is SPGTBs are equivalent to a superposition of 11 symmetric Pearcey Gaussian beams (SPGBs).

## 3 The Pearcey Talbot-like Effect

This section describes the propagation properties of the SPGTBs and CSPGTBs in the free space by simulation and experiment. In previous research [26], for Gaussian beams illumination, the Talbot distance is expressed by $Z_G = 2m\frac{d^2}{\lambda}[1 + (\frac{\lambda z}{\pi w_0})^2]$, where $w_0$ is the Gaussian waist, $d$ is the grating spacing, $z$ is the propagation distance and $m$ is a positive integer. Hence the period is $Z_G = 2m\frac{d^2}{\lambda} * 1.16$, and we approximate it to the Talbot distance of the plane wave for convenience ($Z_G \approx Z_T = 2m\frac{d^2}{\lambda}$). In [4], we demand that $v_n$ is equal to $1/d$. It means that $\triangle$ is equal to $d$. Therefore, the Talbot distance can be expressed as $Z_T = 2m\frac{\triangle^2}{\lambda}$. When $m = 1$, we define that $Z_F = \frac{2\triangle^2}{\lambda} = 0.074Z_R$ ($Z_R$ is the Rayleigh distance $\frac{\pi w_0^2}{\lambda}$) represents one Talbot period, whose position is the primary Talbot image plane and $Z_H = Z_F/2$ represents half of the Talbot period, whose position is the secondary Talbot image plane.

### 3.1 The Symmetric Pearcey Gaussian Talbot-like Beams in the free space

Figure 1 shows the numerical simulation and experimental implementation of the SPGTBs. We have set the SPGTBs to comprise a superposition of 11 SPGBs, shifted along the z-axis with $\triangle$, and each multiplied by the same coefficient 1. Due to the interference between the wavefronts of 11 SPGBs, the intensity distributions of SPGTBs are redistributed in the initial and later planes, manifesting as a matrix distribution of several bright spots, shown in Fig. 1. As shown in Figs. 1(a1)-1(a4), the intensity distributions of each Talbot period are similar, uniform matrix distribution of small bright spots. There are 10, 10, 8, 4 bright spots in each row, and other spots are diverged, corresponding to the Figs. 1(a1)-1(a4), respectively. Because of the propagation property of the SPGBs, the number of the small bright spots decreases as the propagation distance increases, which will explain in Fig. 3. Similarly, the attenuation of the small bright spots on both sides can be seen in the intensity curves [white lines at the bottom of the Figs. 1(a1)-1(a4)]. Although the intensities of the small bright spots around the overall spot at each Talbot period planes decrease as the propagation distance increases, the relative distributions of the small bright spots still have a certain similarity. Therefore, we claim that the SPGTBs have the Talbot-like Effect in the free space. Meanwhile, we have experimentally implemented the SPGTBs through computer-generated hologram (CGH) reproduced in reflective spatial light modulators (rSLM) [21], where the plane wave interferes with the target beams, producing the CGH. Through this method, interference fringes of CGH will load the amplitude and the phase information of the target beams. We emit linearly polarized Gaussian beams (GBs) with a He-Ne laser collimating and expanding it in the experiment. Then the GBs are direct to the rSLM, producing the target beams. At last, the target beams are filtered by a 4f-system and then incident into CCD. Moving the CCD, we can observe the transverse patterns of the target beams in different propagation distances. Figures 1(b1)-1(b3) are the experiment results corresponding to Figs. 1(a1)-1(a3) and the experimental patterns agree with the theoretical ones.

Figure 2 shows the propagation of the SPGBs. From [21], we know that the SPGBs have a long focal plane, and its pattern is a circular main lobe with four side lobes that will split into four off-axis main lobes as the propagation distance increases (about $0.051Z_R$), and finally the lobes will diverge. We observe the transverse patterns of planes 1-7 in Figs. 2(a) and 2(b) with one period, representing fractional Talbot plane [Figs. 2(c2)-2(c8) are equidistant from each other]. The enlarged views of small bright spots of SPGTBs are depicted on the lower left of Figs. 2(c2)-2(c8), which are similar to the SPGBs [Figs. 2(d2)-2(d8) are also equidistant]. The Talbot effect predicts fractional effects in between and so four main lobes around the initial position of each are already expected. Firstly, four main lobes of each beam in the corner slowly focus on the main lobe in the center [Figs. 2(c2)-2(c4)], and then the main lobe splits into four main lobes in the corner again (about $0.028Z_R$) [Figs. 2(c6)-2(c8)]. Different from SPGBs, every single beam of SPGTBs keeps alternating quickly from four main lobes to one main lobe and will not diverge at a certain distance [Fig. 2(a)]. Therefore, we infer that every single beam of SPGTBs will present partial propagation property of SPGBs.

Above is the analysis of every single beam of SPGTBs. For the whole distribution of SPGTBs, the transverse intensities distributions of SPGTBs are also similar to that of the SPGBs during propagation. Figures 2(c1), 2(c5), 2(c9)-2(c12) and figures 2(d1), 2(d5), 2(d9), 2(d12) are also equidistantly from each other, respectively. From Figs. 2(c1), 2(c5), 2(c9), 2(c10), the small bright spots arranged by the matrix have the strongest intensity in the center of the matrix, then the intensity weakens slightly from the center to the outward, and divergent spots are around the matrix, like Figs. 2(d1), 2(d5), 2(d9), 2(d10). Excluding the bright spots caused by the constructive interference between beams, we can observe that the intensity distributions of Figs. 2(c11), 2(c12) and Figs. 2(d11), 2(d12) are similar, according to the distribution of bright spots and divergence spots. Every single beam of SPGTBs itself is a SPGB, so the whole distribution of SPGTBs is similar to SPGBs, the small bright spots around the SPGTBs will attenuate. That is why the number of the small bright spots will decrease [Figs. 1(a1)-1(a4)]. In addition, we know that the SPGTBs have multiple autofocusing property [Figs. 2(a) and 2(b)]. Although every single beam of SPGTBs splits into four off-axis main lobes at every $Z_F$ plane, SPGTBs still present autofocusing because of the constructive interference of the 11 SPGBs. While SPGTBs' autofocusing at every $Z_H$ plane is caused by each beam of SPGTBs' autofocusing and periodic reconstruction of the wavefront by the Talbot Effect.

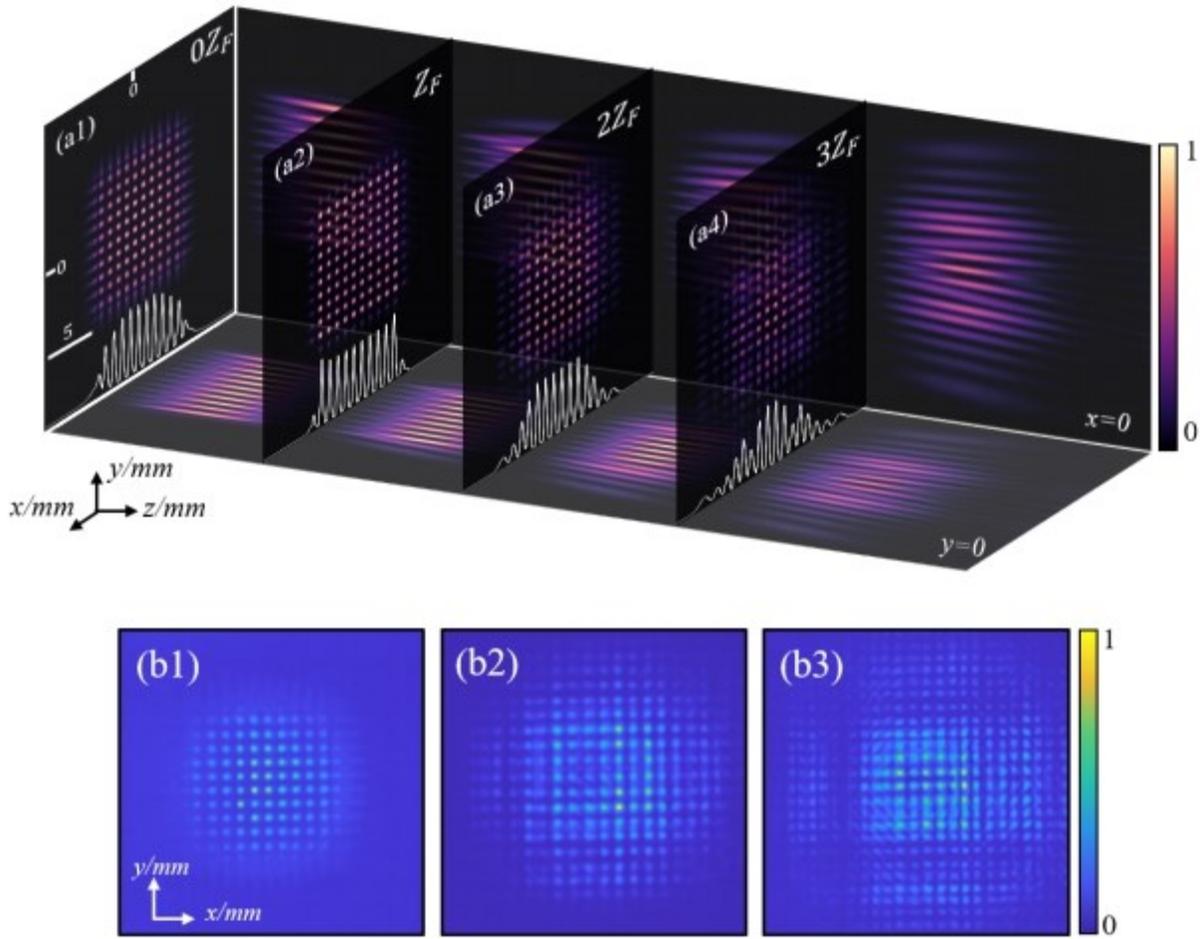

Figure 1: Numerical simulation and experimental implementation of the SPGTBs with $(b, p) = (0.072, 0)$; (a1)-(a4) numerical snapshots of normalized intensity distributions at Talbot period planes each, and the white lines at the bottom of the snapshots are the normalized intensity curves; (b1)-(b3) experimental snapshots corresponding to (a1)-(a3).

### 3.2 The Symmetric Pearcey Gaussian Talbot-like Effect with a cubic phase in the free space

The cubic phase makes the beam have Airy distribution and parabolic trajectory, which helps us observe the transverse intensity distribution of the beam better. We obtain the symmetric cubic Pearcey Gaussian beams (CSPGBs) by imposing the cubic phase in the SPGBs. In the same way, we obtain the CSPGTBs by superimposing CSPGBs. Figure 3 shows the numerical simulation and experimental implementation of the CSPGTBs. Like the SPGTBs, the transverse intensity distributions of CSPGBs is redistributed in the initial and later planes. Analogously, the intensity distributions of the CSPGTBs of each Talbot period are similar [Figs. 3(a1)-3(a4)]. During the propagation, the intensity will be concentrated more on the upper right and radiate to the right and up, whereas the bright spots on the lower and left sides will diverge. As we predetermined, the spots of CSPGTBs emerge Airy distribution because of the cubic phase of CSPGBs. Similar to the SPGTBs, we claim that the CSPGTBs also have the Talbot-like Effect in the free space because the relative distributions of the small bright spots on each Talbot plane still have a certain similarity. The experiment results are shown in Figs. 3(b1)-3(b4), which are in good agreement with Figs. 3(a1)-3(a4).

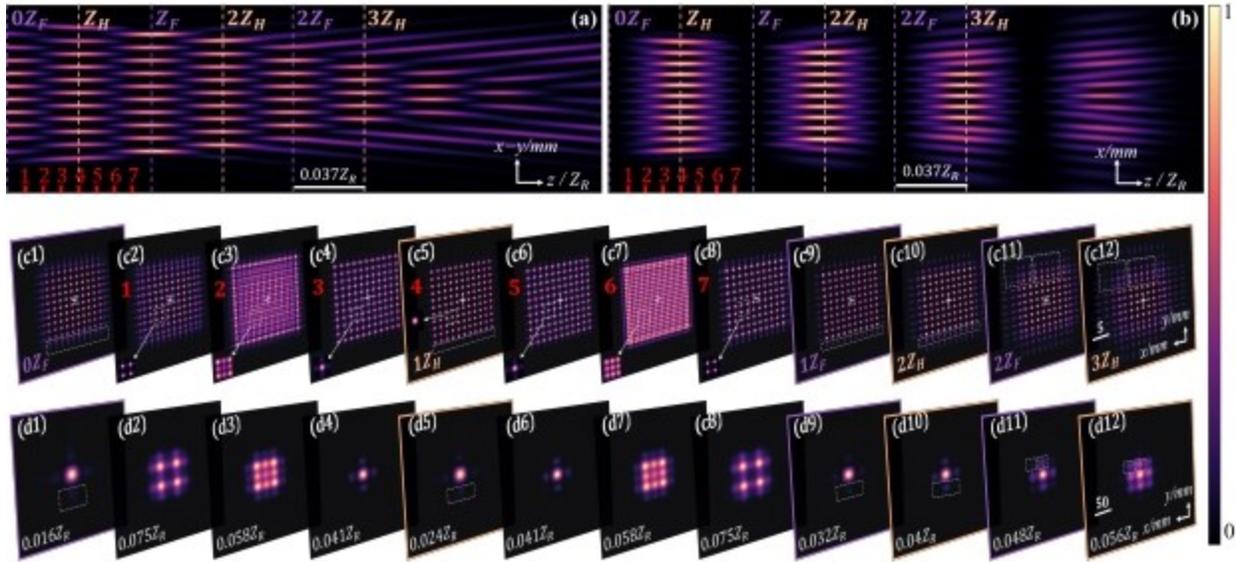

Figure 2: Propagation of the SPGTBs; side view when (a) $x = y$ and (b) $y = 0$; (c1)-(c12) numerical snapshots of normalized intensity distributions of the SPGTBs at planes 1-7 marked in red and at every $Z_H$, $Z_F$ marked in dash lines in (a) and (b), and the white crosses in the center mean the axis origin (0, 0); (d1)-(d12) numerical snapshots of normalized intensity distributions of the SPGBs at different propagation distances. The parameters are the same as those in Fig. 1.

Figure 4 shows the propagation of the CSPGTBs. We know that the CSPGTBs also have multiple autofocusing property [Figs. 4(a) and 4(b)], whose causes are the same as those of SPGTBs. However, the focus plane is not in the Talbot period planes or the secondary Talbot period planes. So the analysis of the period is different from section 3.1. If we want to change the position of focus, we can adjust the distribution factor $b$ or $p$. Besides, we observe the transverse patterns of planes 1-7 in Figs. 4(a) and 4(b) with one period [Figs. 4(d2)-4(d8)], which are equidistant from each other. The enlarged views of small bright spots of CSPGTBs are depicted on the lower left of Figs. 4(d2)-4(d8), which are similar to the CSPGBs [Figs. 4(e2)-4(e8) are also equidistant from each other]. In the same way, the transverse intensity distribution of every single beam of CSPGTBs presents that of CSPGBs. Due to the cubic phase, the transverse intensity distribution of the CSPGTBs isn't symmetric about plane 4 and the propagation trajectory of the CSPGTBs is parabolic. From Figs. 4(d1), 4(d5), 4(d9)-4(d12) which are equidistantly from each other, the small bright spots in the upper right corner have stronger intensity, which is consistent with the case of the CSPGBs [Figs. 4(e1), 4(e5), 4(e9)-4(e12) are equidistant from each other]. The positions of the dash lines in the Figs. 4(e1), 4(e5), 4(e12) have no intensities, and correspondingly, the intensities at the positions of the dash lines in the Figs. 4(d1), 4(d5), 4(d12) are relatively low. As the propagation distance increases, both the CSPGTBs and the CSPGBs present Airy distribution [Figs. 4(d10)-4(d12) and Figs. 4(e10)-4(e12)]. Therefore, the transverse intensity distribution of the CSPGTBs presents that of CSPGBs during propagation. In conclusion, the transverse intensity distributions of every single beam of superimposed beams and the superimposed beams will present that of the single beam, and the superimposed beams have the property of multiply autofocusing (attributed to the focusing of the Talbot-like effect and the autofocusing of Pearcey beams).

## 3.3    Generation of tiny optical bottles by the SPGTBs

It is well known that OBs have a wide range of uses in particle capture. By adjusting the beam shift factor, we can change the interference between each beam and the Talbot period. When setting $\vartriangle = 0.39$mm, we can obtain several tiny OBs shown in Fig. 5(a). From Figs. 5(b1), 5(b3), 5(b5), the transverse pattern of SPGTBs appears as a high-intensity region acting as the top and the bottom of OBs. From Figs. 5(b2) and 5(b4), there are lots of low-intensity cavities distributed evenly in the space acting as the body of OBs. For example, the white circle in Figs. 5(b3),5(b5) is a high-intensity region, a top and bottom of one optical bottle. While the same position in Fig. 5(b4) is a low-intensity cavity surrounded by high-intensity. The position of this optical bottle is depicted by a dotted oval in Fig. 5(a). Therefore, there are many three dimensional dark regions in the free space produced by the SPGTBs, providing many optical traps in three dimensions. The decrease of $\triangle$ causes the spacing between each beam to decrease, and then the intensity distributions are reconstructed. Every single beam of SPGTBs' main lobe is glued together with four main lobes when it splits, forming a high-intensity area (the top and the bottom of OBs). Both traditional OBs and tiny OBs are generated during the propagation of beams. However, traditional OBs are generated through two focuses forming the top and the bottom of the bottles, respectively. In contrast, tiny OBs are generated by the constructive and destructive interference of the Talbot-like effect. Comparing with traditional OBs, although tiny OBs are smaller than traditional OBs, tiny OBs are more in number. Tiny OBs distribute in the form of spatial array in the free space making tiny OBs constraint the particles at various points in space, while traditional OBs constructed by single beam can not. Meanwhile, SPGBs can not form OBs in free space, but with the help of Talbot-like effect, they can. The experiment results are shown in Figs. 5(b1)-5(b5), which are in good agreement with Figs. 5(a1)-5(a5).

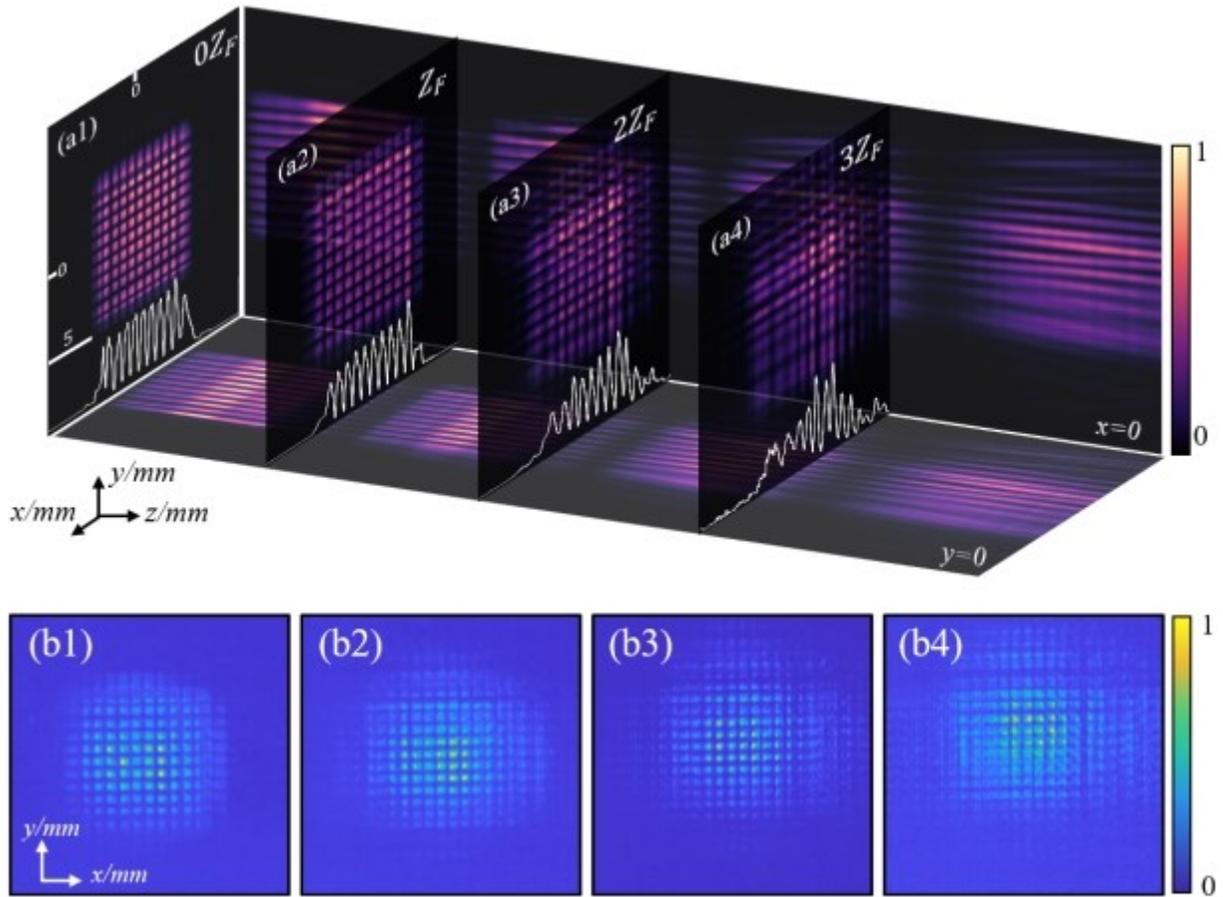

Figure 3: Numerical simulation and experimental implementation of the CSPGTBs with $\beta = 7$, $(b, p) = (0.072, 0)$; (a1)-(a4) numerical snapshots of normalized intensity distributions at Talbot period planes each, and the white lines at the bottom of the snapshots are the normalized intensity curves of Talbot period planes each; (b1)-(b4) experimental snapshots corresponding to (a1)-(a4).

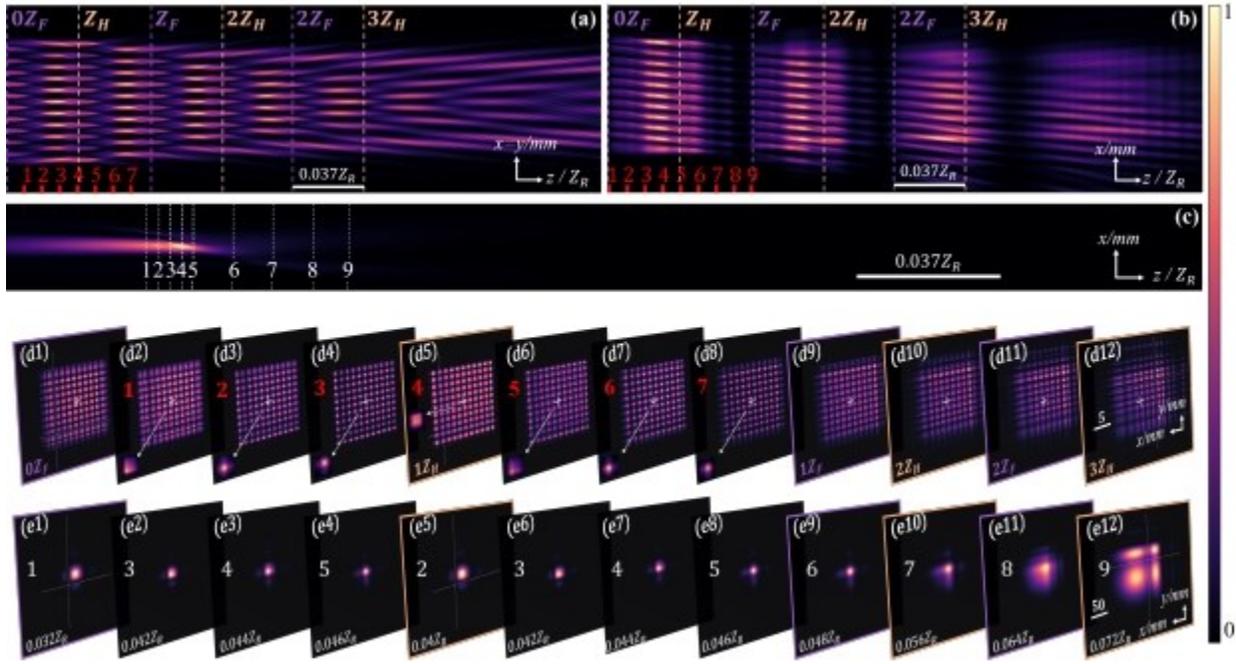

Figure 4: Propagation of the CSPGTBs; side view of the CSPGTBs when (a) $x = y$ and (b) $y = 0$; (c) side view of the CSPGBs when $y = 0$; (d1)-(d12) numerical snapshots of normalized intensity distributions at planes 1-7 marked in red and at every $Z_H$, $Z_F$ marked in dash lines in (a) and (b), and the white crosses in the center mean the axis origin (0, 0); (e1)-(e12) normalized intensity distributions of the CSPGBs at planes 1-9 marked in dash in (c) corresponding to (d1)-(d12). The parameters are the same as those in Fig. 3.

## 4  The influence of parameters on Symmetric Pearcey Gaussian Talbot-like Effect

Inquisitive about the behavior of the symmetric Pearcey Talbot-like effect with different parameters, we observe the SPGTBs under various parameters setting in Fig. 6. Only when we select suitable beam shift factor △ ,distribution factors $b$ and $p$ can the beams show the Talbot-like Effect. Therefore, we select the parameters setting shown in Figs. 6(a)-6(c). The peak intensities in the initial plane and along the z-axis first increase and then decrease with the increase of △. The △ affects not only the focusing positions and times but also the intensity distributions [Figs. 6(a1)-6(a3)]. Firstly, the Talbot period will get longer evenly as △ increasing, which causes the focusing times to reduce and the focusing positions to shift back. Secondly, the intensity is going to be more concentrated in the back if △ gets small. Besides, the increase of the beam shift factor causing periodic interference between each beam, which results in the intensity of the initial plane oscillating up and down instead of change smoothly. Moreover, the overall spot size gets large and the distance between each beam increases with the increase of △ . That is why we can obtain the tiny optical bottles by selecting a small △ in section 3.3.

The changing trends of the normalized peak intensity of the initial plane and along the z-axis of the SPGTBs first increase and then decrease smoothly, as $b$ increasing. The increase of $b$ not only affects the focusing times but also influences the focusing positions. For every single beam, the focal length will be longer with the increase of $b$, and the maximum intensity will shift back [21]. According to the previous conclusion, the superimposed beams will present the single beam's transverse intensity distributions. Therefore, the SPGTBs will have more focuses, and the focal positions will shift back [Figs. 6(b1)-6(b3)], when $b$ increasing. Next, we analyze the influence of the parameter $p$ on the Talbot-like Effect. Unlike △ and $b$, the changing trend of the normalized peak intensity of the initial plane and along the z-axis of the SPGTBs will decrease when $p$ gets larger. Meanwhile, the $p$ also affects the focusing positions and times like $b$. For every single beam, $p$ can also do what the $b$ does. In like manner, the larger $p$ is, the more focusing times are, the farther the focal positions are [Figs. 6(c1)-6(c3)]. Comparing with Figs. 6(a) and 6(b), $p$ has a larger adjustable range than $b$, so $p$ can roughly adjust the focusing positions and times, while $b$ can fine tune the focusing positions and times. However, the larger $p$ will decrease the intensity, so we should choose $b$ and $p$ reasonably. The advantage of adjusting $b$ and $p$ is that we can adjust the focusing positions and times without changing the Talbot period. In addition, the parameters $b$ and $p$ have a fascinating influence on SPGTBs. Under appropriate range, the larger $b$ and $p$ are, the less likely SPGTBs are to diverge and the better the focusing effect is. Comparing three figures in Figs. 6 (b1)-6(b3) and Figs. 6 (c1)-6(c3), respectively, the larger the $b$ and $p$, the more energy is concentrated in the center and farther. In the traditional application of the Talbot effect, a Gaussian beam passing through a finite periodic object, such as a two-dimensional grating, will generate the Gaussian Talbot array [26]. But the array will diverge with the increase of the propagation distance. However, the Talbot-like effect generated by symmetric Pearcey beams in this paper has a focusing effect attributing to $b$ and $p$. In the case of the Talbot effect in lithography, etc, if the position of the image plane (Talbot plane) needs to be changed the only method is changing the period of the grating [26, 31]. Because the periodic structure of the objects decides the position of the Talbot plane. However, we can change the $b$ and $p$ to change the position of the Talbot plane without changing the periodic structure of the objects.

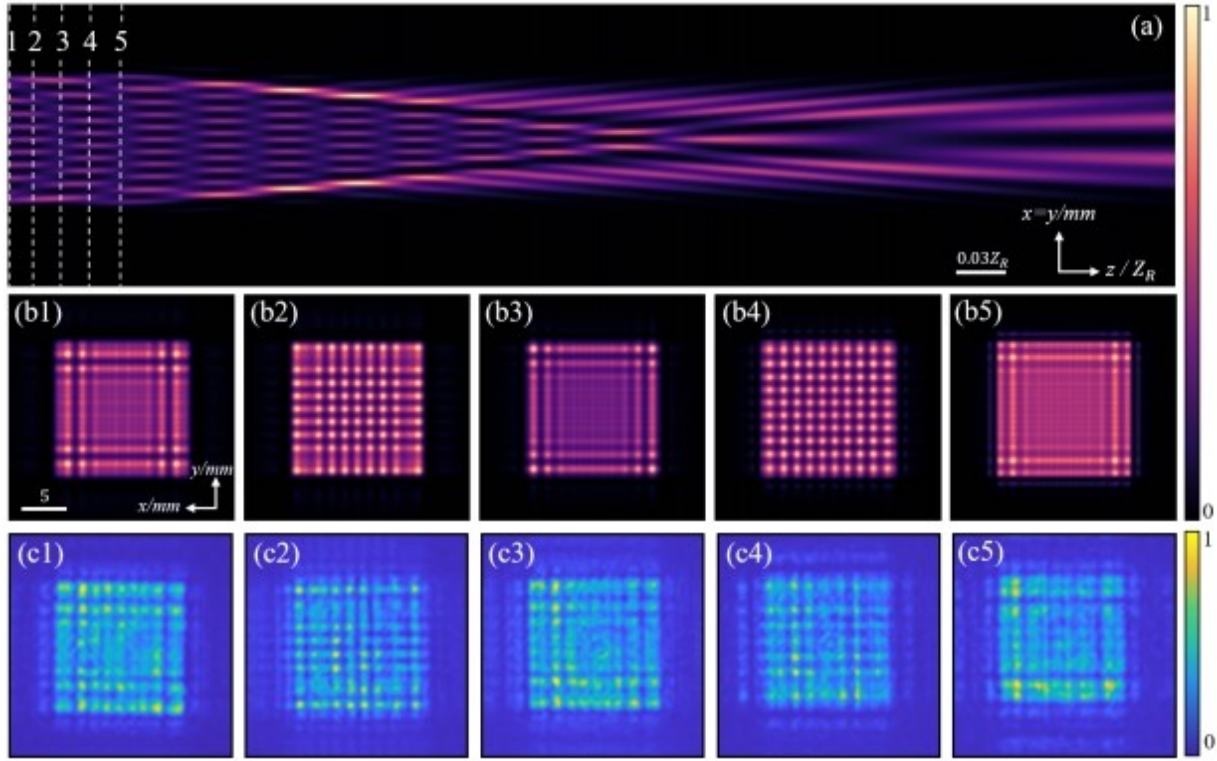

Figure 5: Propagation of SPGTBs with $\vartriangle$ = 3.9; (a) side view of SPGTBs propagation when $x = y$; (b1)-(b5) numerical snapshots of normalized intensity distributions at planes 1-5 marked in (a); (c1)-(c5) experimental snapshots corresponding to (b1)–(b5). Other parameters are the same as those in Fig. 1.

## 5    Conclusion

In conclusion, we have demonstrated the symmetric Pearcey Talbot-like Effect numerically and experimentally by the superposition of the symmetric Pearcey beams. In the symmetric Pearcey Talbot-like effect, the wavefront of the superimposed beams is redistributed periodically because of interference. The transverse intensity distributions of every single beam and the superimposed beams will present that of the single beam. Therefore, the symmetric Pearcey Talbot-like effect can make the symmetric Pearcey beams have the multiple autofocusing property and make the symmetric Pearcey beams form lots of tiny optical bottles under the appropriate parameters setting. Additionally, by adjusting the beam shift factor $\vartriangle$ ,the distribution factors $b$ and $p$, we can also control the intensity distribution, the focusing positions and times of the symmetric Pearcey beams. Our results further connect Talbot effect and symmetric Pearcey beams. This symmetric Pearcey Talbot-like effect not only provides a deeper insight into the self-imaging beams, but also provide some potential applications in Talbot effect's optical applications such as optical imaging, optical traps, lithography, etc.

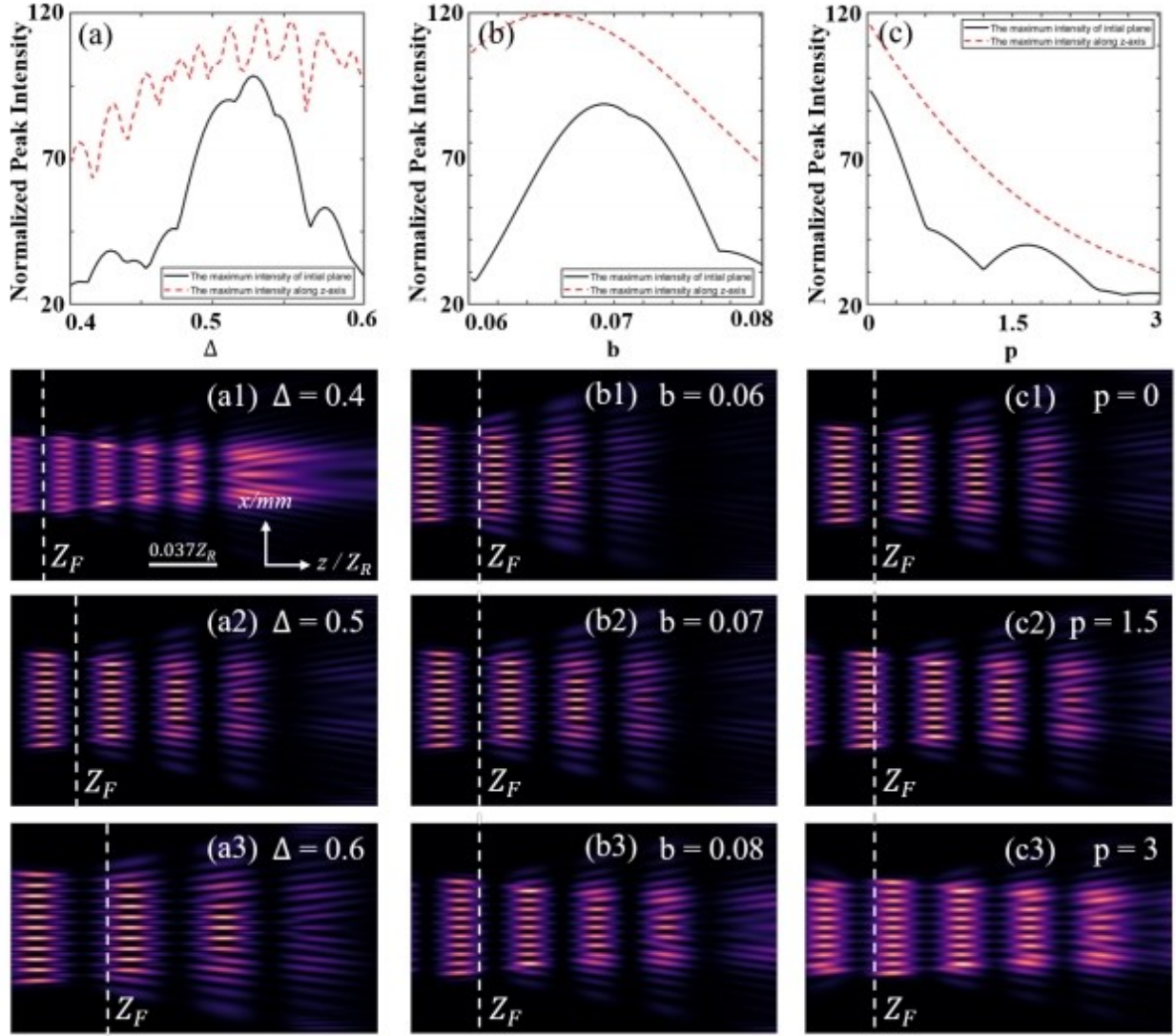

Figure 6: Propagation of SPGTBs under various parameters setting: (a)-(c) Normalized peak intensities of the initial plane and along the z-axis with different $\triangle$, $b$ and $p$, respectively; (a1)-(a3), (b1)-(b3) and (c1)-(c3) side views of SPGTBs when $y = 0$ with $\triangle = 0.4, 0.5, 0.6$, $b = 0.06, 0.07, 0.08$, $p = 0, 1.5, 3$, respectively, and the white dash lines represent the first Talbot period planes. Other parameters are the same as those in Fig. 1.

## Funding.

## Disclosures.

The authors declare no conflicts of interest.